\documentclass[aps,pra,superscriptaddress,longbibliography,floatfix,twocolumn]{revtex4-1}

\usepackage{graphicx}
\usepackage{epstopdf}
\usepackage{amsmath}
\usepackage{amsfonts}
\usepackage{amssymb}
\usepackage{hyperref}
\usepackage{bm}
\usepackage{longtable}

\begin{document}


\title{Families of magnetic semiconductors - an overview}

\author{Tomasz Dietl}
\email{dietl@ifpan.edu.pl}
\affiliation{International Research Centre MagTop, Institute of Physics, Polish Academy of Sciences,
Aleja Lotnikow 32/46, PL-02668 Warsaw, Poland}
\affiliation{WPI Advanced Institute for Materials Research,
Tohoku University, 2-1-1 Katahira, Aoba-ku, Sendai 980-8577, Japan}

\author{Alberta Bonanni}
\email{alberta.bonanni@jku.at}
\affiliation{Institut f\"ur Halbleiter- und Festk\"orperphysik,
Johannes Kepler University, Altenbergerstrasse 69, A-4040 Linz, Austria}

\author{Hideo Ohno}
\email{ohno@riec.tohoku.ac.jp}
\affiliation{Laboratory for Nanoelectronics and Spintronics, Research Institute of Electrical Communication,
Tohoku University, 2-1-1 Katahira, Aoba-ku, Sendai 980-8577, Japan}
\affiliation{Center for Science and Innovation in Spintronics (Core Research Cluster)
Tohoku University 2-1-1 Katahira, Aoba-ku, Sendai 980-8577, Japan}
\affiliation{WPI Advanced Institute for Materials Research,
Tohoku University, 2-1-1 Katahira, Aoba-ku, Sendai 980-8577, Japan}
\affiliation{Center for Spintronics Integrated System,
Tohoku University, 2-1-1 Katahira, Aoba-ku, Sendai 980-8577, Japan}
\affiliation{Center for Innovative Integrated Electronic Systems,
Tohoku University, 468-1 Aramaki Aza Aoba, Aoba-ku, Sendai 980-0845, Japan}
\affiliation{Center for Science and Innovation in Spintronics
(Core Research Cluster),
Tohoku University, 2-1-1 Katahira, Aoba-ku, Sendai 980-8577, Japan}
\affiliation{Center for Spintronics Research Network,
Tohoku University, 2-1-1 Katahira, Aoba-ku, Sendai 980-8577, Japan}


\begin{abstract}
\noindent\textbf{Abstract:}\,\,\, The interplay of magnetic and semiconducting properties has been in the focus since more than a half of the century.
In this introductory article we briefly review the key properties and functionalities of various magnetic semiconductor families, including europium chalcogenides, chromium spinels, dilute magnetic semiconductors, dilute ferromagnetic semiconductors and insulators, mentioning also sources of non-uniformities in the magnetization distribution, accounting for an apparent high Curie temperature ferromagnetism in many systems. Our survey is carried out from today's perspective of ferromagnetic and antiferromagnetic spintronics as well as of the emerging fields of magnetic topological materials and atomically thin 2D layers.

\end{abstract}

\maketitle

\vspace{0.5\baselineskip}

\noindent\textbf{DOI:}\,\,\, \hspace{10mm}\textbf{PACC:}\,\,\,7550P

\subsection*{Magnetic semiconductors}

The discovery of ferromagnetism in some europium chalcogenides and chromium spinels a half of the century ago$^{\mbox{\scriptsize{\cite{Methfessel:1968_B}}}}$ came as a surprise, since insulators typically showed antiferromagnetic or ferrimagnetic spin ordering, driven by a superexchange interaction, whereas ferromagnetism was considered a domain of metals. However, the Goodenough-Kanamori-Anderson rules indicate in which cases the superexchange can lead to ferromagnetic short-range coupling between localized spins. This mechanism accounts e.g. for Curie temperature $T_{\mbox{\scriptsize{C}}} = 130$\,K in CdCr$_2$Se$_4$$^{\mbox{\scriptsize{\cite{Baltzer:1966_PR}}}}$. In the case of EuO and EuS, the antiferromagnetic superexchange is  overcompensated by a direct $f$--$d$ ferromagnetic exchange, which  results in $T_{\mbox{\scriptsize{C}}} = 68$\,K and $16$\,K,  respectively$^{\mbox{\scriptsize{\cite{Wachter:1979_HB}}}}$.

Soon after their discovery, magnetic semiconductors were found to exhibit outstanding properties, including colossal magnetoresistance and magnetooptical effects, assigned rightly to the exchange interaction of band carriers and localized spins. This $sd$--$f$ coupling results in a giant spin splitting of bands below $T_{\mbox{\scriptsize{C}}}$, as well as in magnetization fluctuations near $T_{\mbox{\scriptsize{C}}}$ generated by band carriers$^{\mbox{\scriptsize{\cite{Kasuya:1968_RMP,Nagaev:2001_PR}}}}$. Due to the presence of a sizable $sd$--$f$ exchange interaction, by electron doping (oxygen vacancies, Eu substituted by Gd) the magnitude of $T_{\mbox{\scriptsize{C}}}$ can be enhanced by about 50\,K in EuO, in agreement with the Ruderman-Kittel-Kasuya-Yosida (RKKY) theory$^{\mbox{\scriptsize{\cite{Wachter:1979_HB}}}}$.

More recently, EuS and related ferromagnetic insulators have been proposed as functional overlayers that can lead to novel topological properties by exchange splitting of interfacial bands {\em{via}} a ferromagnetic proximity effect$^{\mbox{\scriptsize{\cite{Zutic:2019_MT}}}}$. At the same time, HgCr$_2$Se$_4$ ($T_{\mbox{\scriptsize{C}}} = 110$\,K) appears to be a Weyl semimetal$^{\mbox{\scriptsize{\cite{Xu:2011_PRL}}}}$. Furthermore, antiferromagnetic EuTe constitutes a test-bench for the emerging field of antiferromagnetic spintronics$^{\mbox{\scriptsize{\cite{Jungwirth:2016_NN}}}}$.

Another class of materials that attract considerable attention are van der Waals  magnetic semiconductors studied down to the atomically thin limit$^{\mbox{\scriptsize{\cite{Li:2019_AM}}}}$. A competition between ferromagnetic and antiferromagnetic superexchange accounts for the magnetic properties of CrI$_3$ and related systems  as a function of the number of layers, electric field, and strain$^{\mbox{\scriptsize{\cite{Li:2019_AM,Webster:2018_PRB}}}}$. In particular, monolayers are ferromagnetic in the case of these compounds in which, similarly to 2D (Cd,Mn)Te$^{\mbox{\scriptsize{\cite{Haury:1997_PRL}}}}$, the low-temperature 2D spin order is stabilized by the large magnitude of the uniaxial magnetic anisotropy$^{\mbox{\scriptsize{\cite{Huang:2017_N}}}}$. On the other hand, TMPX$_3$, where TM is a transition metal and X is either S or Se, are antiferromagnetic semiconductors. Their properties as a function of the layer number are under investigations$^{\mbox{\scriptsize{\cite{Li:2019_AM}}}}$.

 \subsection*{Dilute magnetic semiconductors}
 This material family was initially named {\em semimagnetic semiconductors}$^{\mbox{\scriptsize{\cite{Galazka:1978_P}}}}$, as it comprises standard semiconductors doped with magnetic impurities that are randomly distributed, electrically inactive, and do not show any long-range spatial spin ordering$^{\mbox{\scriptsize{\cite{Furdyna:1988_B,Dietl:1994_B}}}}$. Typical representatives are (Cd,Mn)Te, (Zn,Co)O, and (Pb,Eu)S. Here, the short-range antiferromagnetic  superexchange between Mn$^{\mbox{\scriptsize{\cite{Spalek:1986_PRB,Galazka:2006_pssb}}}}$ and heavier 3$d$ transition metal ions (e.g. Co) or long-range dipolar interactions between rare earth magnetic moments$^{\mbox{\scriptsize{\cite{Andresen:2014_PRX}}}}$ lead to spin-glass freezing at $T_{\mbox{\scriptsize{f}}} < 1$\,K for $x < 0.1$.  Magnetooptical and quantum transport experiments have allowed to reveal and quantify the influence of $sp$--$d$ coupling upon the exciton, polariton, and Landau level spectra, quantum Hall effects, one-electron and many-body quantum localization, universal conductance fluctuations, and carrier spin dynamics in dilute magnetic semiconductor systems of various dimensionalities realized in bulk, 2D quantum structures, nanowires, and quantum dots$^{\mbox{\scriptsize{\cite{Dietl:1994_B,Gaj:2010_B,Krol:2018_SR,Betthausen:2014_PRB}}}}$, as well as made it possible to optically detect electrical spin injection$^{\mbox{\scriptsize{\cite{Fiederling:1999_N}}}}$. Conversely, the dynamics of localized spins has been probed by the Faraday effect$^{\mbox{\scriptsize{\cite{Leclercq:1993_PRB}}}}$ or quantum noise spectroscopy$^{\mbox{\scriptsize{\cite{Jaroszynski:1998_PRL}}}}$. The physics of bound magnetic polarons, single electrons interacting with spins localized within the confining potential of shallow impurities or quantum dots, has been advanced in dilute magnetic semiconductors$^{\mbox{\scriptsize{\cite{Dietl:2015_PRB}}}}$. While the majority of experimental results can be explained within virtual crystal and molecular field approximations, strong coupling effects show up in the case of oxides and nitrides doped with transition metal impurities$^{\mbox{\scriptsize{\cite{Dietl:2008_PRB,Pacuski:2008_PRL}}}}$.

 Most of the end binary compounds, for instance MnSe and EuTe, are insulating antiferromagnets and, therefore, attract attention from the point of view of antiferromagnetic spintronics. Another ultimate limit of dilute magnetic semiconductors is represented by qubit systems consisting of single magnetic ions in single quantum dots$^{\mbox{\scriptsize{\cite{Besombes:2004_PRL,Kobak:2014_NC}}}}$.

\subsection*{p-type dilute ferromagnetic semiconductors}
 In these systems, a high density of delocalized or weakly localized holes leads to long-range ferromagnetic interactions between transition metal cations, which dominate over the antiferromagnetic superexchange, and are well described by the $p$--$d$ Zener model$^{\mbox{\scriptsize{\cite{Dietl:2000_S,Dietl:2014_RMP}}}}$, more universal compared to RKKY-type of approaches.  The flagship example of this material family is (Ga,Mn)As in which Mn ions introduce spins and holes to the valence band$^{\mbox{\scriptsize{\cite{Ohno:1998_S}}}}$, but to this category belong also other types of magnetically doped p-type compounds, in which holes originate from point defects, like (Pb,Sn,Mn)Te $^{\mbox{\scriptsize{\cite{Story:1986_PRL}}}}$, or from shallow acceptor impurities, {\em e.g.}, (Cd,Mn)Te/(Cd,Mg)Te:N$^{\mbox{\scriptsize{\cite{Haury:1997_PRL}}}}$ and (Zn,Mn)Te:N$^{\mbox{\scriptsize{\cite{Ferrand:2000_JCG}}}}$, rather than from Mn ions.  The reported magnitudes of Curie temperature $T_{\mbox{\scriptsize{C}}}$ reach 200\,K in (Ga,Mn)As$^{\mbox{\scriptsize{\cite{Olejnik:2008_PRB,Wang:2008_APL,Chen:2011_NL}}}}$, (Ge,Mn)Te$^{\mbox{\scriptsize{\cite{Fukuma:2008_APLa,Hassan:2011_JCG}}}}$, and  (K,Ba)(Zn,Mn)$_2$As$_2$$^{\mbox{\scriptsize{\cite{Zhao:2014_ChSB}}}}$ with less than 10\% of Mn cations, $x_{\mbox{\scriptsize{eff}}} < 0.1$, measured by saturation magnetization in moderate fields, $\mu_0H \lesssim 5$\,T. The values of Curie temperatures achieved in p-type dilute ferromagnetic semiconductors are to be contrasted with $T_{\mbox{\scriptsize{C}}} \approx 0.16$\,K in n-Zn$_{1-x}$Mn$_x$O:Al with $x = 0.03$$^{\mbox{\scriptsize{\cite{Andrearczyk:2001_ICPS}}}}$. Such low values of $T_{\mbox{\scriptsize{C}}}$ in n-type systems reflect the relatively small magnitudes of both $s$--$d$ exchange integral and density of states. Higher $T_{\mbox{\scriptsize{C}}}$ values are observed only in specific situations, for instance, at the crossings of electron Landau levels under quantum Hall effect conditions$^{\mbox{\scriptsize{\cite{Kazakov:2017_PRL}}}}$.

 Ground breaking spintronic functionalities have been demonstrated and theoretically described for (Ga,Mn)As and related systems$^{\mbox{\scriptsize{\cite{Dietl:2014_RMP,Jungwirth:2014_RMP}}}}$.  They rely on the strong $p$--$d$ coupling between localized spins and hole carriers, as well as on sizable spin-orbit interactions in $p$-like orbitals forming the valence band or originating from inversion asymmetry of the host crystal structure. Numerous functionalities of (Ga,Mn)As and other p-type dilute ferromagnetic semiconductors (electrical spin injection$^{\mbox{\scriptsize{\cite{Ohno:1999_N}}}}$, magnetization control by an electric field$^{\mbox{\scriptsize{\cite{Ohno:2000_N,Boukari:2002_PRL,Chiba:2008_N}}}}$, current-induced domain-wall motion$^{\mbox{\scriptsize{\cite{Yamanouchi:2004_N,Yamanouchi:2006_PRL}}}}$, anisotropic tunneling magnetoresistance$^{\mbox{\scriptsize{\cite{Gould:2004_PRL,Wunderlich:2006_PRL}}}}$, and spin-orbit torque$^{\mbox{\scriptsize{\cite{Bernevig:2005_PRB,Chernyshov:2009_NP}}}}$) have triggered the spread of spintronic research over virtually all materials families. In particular, they are now explored in multilayers of transition metals supporting ferromagnetism above room temperature$^{\mbox{\scriptsize{\cite{Kanai:2015_AAPPSBL}}}}$. At the same time, the search for high $T_{\mbox{\scriptsize{C}}}$ ferromagnetic semiconductors led to the discovery of Fe-based superconductors, whereas theoretical approaches to the anomalous Hall effect in (Ga,Mn)As in terms of the Berry phase$^{\mbox{\scriptsize{\cite{Jungwirth:2002_PRL,Nagaosa:2010_RMP}}}}$ and studies of spin-orbit torque in the same compound$^{\mbox{\scriptsize{\cite{Bernevig:2005_PRB,Chernyshov:2009_NP}}}}$ have paved the way to spintronics of topological$^{\mbox{\scriptsize{\cite{Ke:2018_ARCMP,Tokura:2019_NRP}}}}$ and antiferromagnetic systems$^{\mbox{\scriptsize{\cite{Manchon:2019_RMP}}}}$.

\subsection*{Dilute ferromagnetic insulators}
 As mentioned, the Goodenough-Kanamori-Anderson rules indicates in which cases superexchange can lead to ferromagnetic short-range coupling between localized spins.  According to experimental and theoretical studies, such a mechanism operates for Mn$^{3+}$ ions in GaN and accounts for $T_{\mbox{\scriptsize{C}}}$ values reaching about 13\,K at $x \approx 10$\% in semi-insulating wurtzite Ga$_{1-x}$Mn$_x$N $^{\mbox{\scriptsize{\cite{Stefanowicz:2013_PRB}}}}$, an accomplishment preceded by a long series of material development stages$^{\mbox{\scriptsize{\cite{Bonanni:2011_PRB,Kunert:2012_APL}}}}$. Since there is no competing antiferromagnetic interactions and due to high atom density, an unusually high magnitude of magnetization for the dilution $x = 0.1$ is observed in Ga$_{1-x}$Mn$_{x}$N$^{\mbox{\scriptsize{\cite{Kunert:2012_APL}}}}$. In contrast to random antiferromagnets (such as II-VI dilute magnetic semiconductors), there is no frustration in the case of ferromagnetic spin-spin interactions. Furthermore, according to the tight-binding theory, the interaction energy decays exponentially with the spin-spin distance$^{\mbox{\scriptsize{\cite{Stefanowicz:2013_PRB}}}}$. Accordingly, the dependence of $T_{\mbox{\scriptsize{C}}}$ on $x$ cannot be explained within the mean-filed approximation, but corroborates the percolation theory$^{\mbox{\scriptsize{\cite{Korenblit:1973_PLA}}}}$, and can be quantitatively described by combining tight-binding theory and Monte Carlo simulations$^{\mbox{\scriptsize{\cite{Stefanowicz:2013_PRB}}}}$.

The significant inversion symmetry breaking specific to wurtzite semiconductors and the insulating character of the system allow controlling the magnetization by an electric field {\em{via}} a piezoelectromagnetic coupling in (Ga,Mn)N. As shown experimentally and confirmed theoretically$^{\mbox{\scriptsize{\cite{Sztenkiel:2016_NC}}}}$, the inverse piezoelectric effect changes the magnitude of the single-ion magnetic anisotropy specific to Mn$^{3+}$ ions in GaN, and thus the magnitude of magnetization.

\subsection*{Dilute magnetic topological materials}
 In these systems$^{\mbox{\scriptsize{\cite{Ke:2018_ARCMP,Tokura:2019_NRP}}}}$, there are novel consequences of magnetization-induced giant $p$--$d$ exchange spin-splittings of bulk and topological boundary states. In particular, in thin layers of 3D topological insulators, this splitting turns hybridized topological surface states into chiral edge states. Striking consequences of this transformation include: (i) the precise quantization of the Hall conductance $\sigma_{xy} = e^2/h$ demonstrated for thin layers of ferromagnetic (Bi,Sb,Cr)$_2$Te$_3$ at millikelvin temperatures$^{\mbox{\scriptsize{\cite{Chang:2013_S}}}}$, as predicted theoretically$^{\mbox{\scriptsize{\cite{Yu:2010_S}}}}$; (ii)  the efficient  magnetization switching by spin-locked electric currents in these ferromagnets$^{\mbox{\scriptsize{\cite{Fan:2016_NN}}}}$. In the case of 3D Dirac materials with magnetic impurities coupled by antiferromagnetic superexchange, such as (Cd,Mn)$_3$As$_2$ or strained (Hg,Mn)Te, the magnetic field-induced formation of Weyl semimetals with non-zero topological charges has been predicted  $^{\mbox{\scriptsize{\cite{Bulmash:2014_PRB}}}}$.

 Furthermore, inverted band ordering specific to topological matter enhances the role of the long-range interband Bloembergen-Rowland contribution to spin-spin  interactions, resulting in higher spin-glass freezing temperatures $T_{\mbox{\scriptsize{f}}}$ in topological semimetals, such as (Cd,Mn)$_3$As$_2$ and (Hg,Mn)Te,  compared to  topologically trivial II-VI Mn-based dilute magnetic semiconductors$^{\mbox{\scriptsize{\cite{Galazka:2006_pssb}}}}$. This spin-spin exchange mechanism, taken into consideration within the $p$--$d$ Zener model$^{\mbox{\scriptsize{\cite{Dietl:2001_PRB}}}}$ and named Van Vleck's paramagnetism, was proposed to lead to ferromagnetism in topological insulators of bismuth/antimony chalcogenides containing V, Cr, or Fe$^{\mbox{\scriptsize{\cite{Yu:2010_S}}}}$, a proposal discussed earlier in the context of zero-gap (Hg,Mn)Te$^{\mbox{\scriptsize{\cite{Lewiner:1980_JPC}}}}$. However, according to {\em ab initio} studies and corresponding experimental data, the superexchange in the case of V and Cr, and the RKKY coupling in Mn-doped films appear to contribute significantly to ferromagnetism in tetradymite V$_2$-VI$_3$ compounds$^{\mbox{\scriptsize{\cite{Vergniory:2014_PRB}}}}$. It appears that the co-existence of ferromagnetic superexchange with ferromagnetic Bloembergen-Rowland and RKKY interactions accounts for $T_{\mbox{\scriptsize{C}}}$ as high as 250\,K in (Cr$_x$Sb$_{1- x}$)$_2$Te$_3$ with $x = 0.44$$^{\mbox{\scriptsize{\cite{Gupta:2017_APEX}}}}$. Ferromagnetic coupling between magnetic impurities mediated by carriers residing in surface or edge Dirac cones is rather weak for realistic values of spatial extension of these states into the bulk$^{\mbox{\scriptsize{\cite{Dietl:2014_RMP}}}}$.

\subsection*{Heterogeneities in magnetic semiconductors}
\label{sec:heterogenous}
It becomes increasingly clear that spatial inhomogeneities in the magnetization distribution account for a number of key properties of magnetic semiconductors, including the magnitude of $T_{\mbox{\scriptsize{C}}}$. We discuss here three distinct mechanisms accounting for the non-uniform distribution of magnetization and describe their consequences.

(1) According to the percolation theory applicable to dilute systems with short-range spin-spin interactions, critical temperature $T_c$ corresponds to the formation of a percolative cluster that incorporates typically about 16\% of the spins in the 3D case.  This cluster extends over all spins at $T =0$, implying that at $T > 0$ ferromagnetism co-exists with superparamagnetism produced by not percolative clusters$^{\mbox{\scriptsize{\cite{Lachmane:2015_SA}}}}$. Presumably this effect accounts for thermal instabilities of the quantum anomalous Hall effect even at $T \ll T_{\mbox{\scriptsize{C}}}$, and shifts  below 100 mK the operation range of the potential resistance standards based on this phenomenon$^{\mbox{\scriptsize{\cite{Goetz:2018_APL,Fox:2018_PRB}}}}$.

(2) A specific feature of conducting magnetic semiconductors is the interplay of carrier-mediated ferromagnetism with carrier localization, which results in spatial fluctuations of magnetization and superparamagnetic signatures generated by the critical fluctuations in the carrier density of states in the vicinity  of the metal-to-insulator transition. This effect has been extensively discussed in the context of both  ferromagnetic$^{\mbox{\scriptsize{\cite{Nagaev:2001_PR,Pohlit:2018_PRL}}}}$ and dilute ferromagnetic semiconductors$^{\mbox{\scriptsize{\cite{Dietl:2008_JPSJ,Sawicki:2010_NP,Richardella:2010_S}}}}$.

(3) The magnetic properties discussed above have concerned systems with a random distribution of cation-substitutional magnetic impurities. A major challenge in dilute magnetic materials is the crucial dependence of their properties on the spatial distribution of magnetic ions and their position in the crystal lattice. These nanoscale structural characteristics depend, in turn, on the growth and processing protocols, as well as on doping with shallow impurities$^{\mbox{\scriptsize{\cite{Kuroda:2007_NM,Bonanni:2008_PRL}}}}$. By employing a range of photon, electron, and particle beam  methods, with structural, chemical, and spin resolution down to the nanoscale, it has become possible to correlate the surprising magnetic properties with the spatial arrangement and with the electronic configuration of the magnetic constituent$^{\mbox{\scriptsize{\cite{Bonanni:2011_SST,Dietl:2015_RMP}}}}$. In particular, the aggregation of magnetic cations which are either introduced deliberately or present due to contamination,  accounts for the high  $T_{\mbox{\scriptsize{C}}}$ values observed in a number of semiconductors and oxides$^{\mbox{\scriptsize{\cite{Sato:2010_RMP,Bonanni:2009_RCS}}}}$. The symmetry lowering associated with this aggregation is also responsible for remarkable magnetic and magnetotransport anisotropy properties of (Ga,Mn)As$^{\mbox{\scriptsize{\cite{Birowska:2012_PRL}}}}$ and (In,Fe)As$^{\mbox{\scriptsize{\cite{Yuan:2018_PRM}}}}$.

\subsection*{Outlook}
Semiconductors and insulators doped with magnetic impurities, in addition
to showing fascinating physics, have well established applications as semi-insulating substrates and layers, and as building-blocks for
solid state lasers, optical insulators,  and detectors of high energy photons. More recent works have revealed surprising materials science features, such as a dependence of the spatial distribution of magnetic impurities on the Fermi level position.
At the same time, studies of dilute ferromagnetic semiconductors have led
to discoveries of breakthrough functionalities, like for instance
spin-orbit torque and  magnetization manipulation by an electric field,
which are on the way of recasting the concept of computer hardware. The rise of
magnetic quantum materials opens for magnetic
semiconductors a new chapter  involving magnetic proximity effects,
functionalities of topological boundary states, and the control of
 van der Waals heterostructures.  It is justified to expect
 that research in semiconductors, insulators, and
organic materials containing magnetic constituents, and particularly in their multilayers and quantum structures,  will continue  to generate
unanticipated and inspiring discoveries in the years to come.

\section*{Acknowledgements}

The work of T.\,D. is supported by the Foundation for Polish Science through the
IRA Programme financed by EU within SG OP Programme. A.\,B. acknowledges a support by the Austrian Science Foundation --
FWF (P31423 and P26830) and by the Austrian Exchange Service (\"OAD) project PL-01/2017.

\end{document}